\documentclass[10pt,conference]{IEEEtran}
\IEEEoverridecommandlockouts
\usepackage{cite}
\usepackage{amsmath,amssymb,amsfonts}
\usepackage{algorithmic}
\usepackage{graphicx}
\usepackage{textcomp}
\usepackage[table,xcdraw]{xcolor}

\usepackage[citecolor=black,
            colorlinks=true,
            linkcolor=black,
            urlcolor  = black,
            pdfpagelabels={true},
            pdftitle={On Data Sampling Strategies for Training Neural Network Speech Separation Models}
            pdfauthor={William Ravenscroft, Stefan Goetze, Thomas Hain},
            pdfkeywords={speech separation, context modelling, data sampling, speech enhancement, transformer},
            pdfsubject={speech separation, context modelling, convolutional neural networks, speech enhancement, transformer}]
            {hyperref}
\usepackage[english]{babel}
\addto\extrasenglish{}
\addto\extrasenglish{}
\addto\extrasenglish{}
\addto\extrasenglish{}
\addto\extrasenglish{}
\addto\extrasenglish{}
\usepackage{acronym} 
\usepackage{svg}
\usepackage{url}
\usepackage{orcidlink}

\acrodef{AC}    {acoustic conditions}
\acrodef{ASR}   {automatic speech recognition}
\acrodef{BLSTM} {bidirectional long short term memory}
\acrodef{BS}    {batch size}
\acrodef{CB}    {convolutional block}
\acrodefplural{CB}[CBs]{convolutional blocks}
\acrodef{Conv-TasNet} {convolutional \ac{TasNet}}
\acrodef{conformer}{convolution-augmented Transformer}
\acrodef{CSM}   {clean speech mixtures}
\acrodef{cLN}   {channelwise layer normalization}
\acrodef{DAE}   {denoising autoencoder}
\acrodef{DE}    {density estimation}
\acrodef{D-Conv}{depthwise convolution}
\acrodef{DD-Conv}{deformable depthwise convolution}
\acrodef{DDS-Conv}{deformable depthwise-separable convolution}
\acrodef{DS-Conv}{depthwise-separable convolution}
\acrodef{DM}    {dynamic mixing}
\acrodef{DL}    {deep learning}
\acrodef{DNN}   {deep neural network}
\acrodef{DPRNN} {dual path recurrent neural network model}
\acrodef{DPTNet}{dual path Transformer network}
\acrodef{DTCN}  {deformable temporal convolutional network}
\acrodef{ED}    {epoch duration}
\acrodef{E2E}   {end-to-end}
\acrodef{ER}    {early reflection}
\acrodef{gLN}   {global layer normalization}
\acrodef{GRU}   {gated recurrent unit}
\acrodef{GC}    {global context}
\acrodef{FLOP/s} {floating point operations per second}
\acrodef{LSTM}  {long short term memory}
\acrodef{LR}    {late reflection}
\acrodefplural{LR}[LRs]{late reflections}
\acrodef{LUFS}   {loudness units relative to full scale}
\acrodef{MACs}  {mutiply-acccumulate operations}
\acrodef{MHA}   {Multihead Attention}
\acrodef{MOS}   {Mean Opinion Score}
\acrodef{MR}    {mask refinement}
\acrodef{NMF}   {non-negative matrix factorization}
\acrodef{NSM}   {noisy speech mixture}
\acrodef{NRSM}  {noisy reverberant speech mixture}
\acrodef{PESQ}  {perceptual evaluation of speech quality}
\acrodef{PIT}   {permutation invariant training}
\acrodef{PM}    {post-masking}
\acrodef{P-Conv}{pointwise convolution}
\acrodef{PReLU} {parametric \ac{ReLU}}
\acrodef{Q1}    {1st quartile}
\acrodef{Q2}    {2nd quartile}
\acrodef{Q3}    {3rd quartile}
\acrodef{Q4}    {4th quartile}
\acrodef{ReLU}  {rectified linear unit}
\acrodef{RF}    {receptive field}
\acrodef{RIR}   {room impulse response}
\acrodef{RNN}   {recurrent neural network}
\acrodef{RSM}   {reverberant speech mixture}
\acrodef{SA}    {speed augmentation}
\acrodef{SC}    {skip connection}
\acrodef{SepFormer}{separation transformer}
\acrodef{SISDR} {scale-invariant signal-to-distortion ratio}
\acrodef{SDR}   {signal-to-distortion ratio}
\acrodef{SNR}   {signal-to-noise ratio}
\acrodef{SRMR}  {speech-to-reverberation modulation energy ratio}
\acrodef{SSR}   {speech-to-speech ratio}
\acrodef{SP}    {signal processing}
\acrodef{STFT}  {short-time Fourier transform}
\acrodef{STOI}  {short-time objective intelligibility}
\acrodef{SOTA}  {state-of-the-art}
\acrodef{SW}    {shared weights}
\acrodef{ESTOI} {extended short-time objective intelligibility}
\acrodef{TasNet}{time-domain audio separation network}
\acrodef{TCN}   {temporal convolutional network}
\acrodef{TSL}   {training signal length}
\acrodef{UPGMA} {unweighted pair group method with arithmetic mean}
\acrodef{WER}   {word error rate}
\acrodef{WPE}   {weighted prediction error}

\newcommand{\vek}[1]{\ensuremath{\mathbf{#1}}}    
\newcommand{\vekt}[1]{\ensuremath{\boldsymbol{\mathrm{#1}}}}

\newcommand{\Real}{\mathbb{R}}

\def\BibTeX{{\rm B\kern-.05em{\sc i\kern-.025em b}\kern-.08em
    T\kern-.1667em\lower.7ex\hbox{E}\kern-.125emX}}
\begin{document}

\title{
On Data Sampling Strategies for Training Neural Network Speech Separation Models\\
\thanks{{This work was supported by the Centre for Doctoral Training in Speech and Language Technologies (SLT) and their Applications funded by UK Research and Innovation [grant number EP/S023062/1].} {This work was also funded in part by 3M Health Information Systems, Inc.}}
}

\author{\IEEEauthorblockN{William Ravenscroft$^{\orcidlink{0000-0002-0780-3303}}$, Stefan Goetze$^{\orcidlink{0000-0003-1044-7343}}$, and Thomas Hain$^{\orcidlink{0000-0003-0939-3464}}$}

\IEEEauthorblockA{\textit{Department of Computer Science}, 
\textit{{The} University of Sheffield}, Sheffield, United Kingdom \\
\{jwravenscroft1, s.goetze, t.hain\}@sheffield.ac.uk\vspace*{-0.2cm}}
}
\maketitle

\begin{abstract}
Speech separation remains an important area of multi-speaker signal processing. \Ac{DNN} models have attained the best performance on many speech separation benchmarks. Some of these models can take significant time to train and have high memory requirements.
Previous work has proposed shortening training examples to address these issues but the impact of this on model performance is not yet well understood. In this work, the impact of applying these \ac{TSL} limits is analysed for two speech separation models: SepFormer, a transformer model, and Conv-TasNet, a convolutional model. The WJS0-2Mix, WHAMR and Libri2Mix datasets are analysed in terms of signal length distribution and its impact on training efficiency. It is demonstrated that, for specific distributions, applying specific \ac{TSL} limits results in better performance. This is shown to be mainly due to randomly sampling the start index of the waveforms resulting in more unique examples for training. A SepFormer model trained using a \ac{TSL} limit of 4.42s and \ac{DM} is shown to match the best-performing SepFormer model trained with \ac{DM} and unlimited signal lengths. Furthermore, the 4.42s \ac{TSL} limit results in a 44\% reduction in training time with WHAMR.
\end{abstract}


\section{Introduction}
\label{sec:intro}
Speech separation models are used in a number of downstream speech processing tasks, from meeting transcription to assistive hearing \cite{FFASRHaebUmbach,Moritz2017,Raj2021IntegrationOS,ssha}. Often, speakers are at a far-field distance from the microphone, which creates additional challenges for speech separation due to interference from noise and reverberation in the signal \cite{chime5, IWAENCbestpaper, atttasnet}. 

\Ac{DNN} separation models have led to significant improvements on anechoic data but there is a performance gap when these models are used for more distorted speech data \cite{Ditter2019, sepformer,QDPN}. Many of these models use Transformer or \ac{BLSTM} layers \cite{dprnn, sepformer, QDPN} which can consume large amounts of memory and have quadratic time-complexity, i.e. for $L$ input frames of data the model performs at least $L^2$ operations \cite{Katharopoulos2020}. This is a particular concern in training when memory requirements are higher due to storing gradients for each operation required in the back-propagation stage \cite{haykin2009neural}. This increased computational load also means longer training times. One way to compensate for the memory requirements is to use a batch size of 1 \cite{sepformer} which leads to even longer training times as more parameter updates are performed. 
Another approach that reduces memory requirements and allows for larger batch sizes is to reduce the mixture signal length \cite{convtasnet}. 
This reduces the training time but potentially at the expense of performance.

In this work, the first aim is to address if there are \ac{TSL} limits at which no additional performance gain can be attained by \ac{DNN} speech separation models. It is shown that, depending on model and dataset selection, there is a \ac{TSL} limit at which not only no additional performance gain is attained but actually limiting the \ac{TSL} to a specific value can lead to notably improved performance. This effect is demonstrated to be due to a random sampling of the start index when using \ac{TSL} limits. Further evaluations show the benefit of having more unique training examples than using the full signal lengths. Finally, the application of \ac{TSL} limits used with \ac{DM} \cite{wavesplit} is evaluated.

The remainder of this paper is structured as follows. In \autoref{sec:sigmod}, the signal model is introduced. The separation networks and datasets used are described in \autoref{sec:sepmodels}, and \autoref{sec:data}, respectively.  \autoref{sec:evaluations} presents evaluations for varying \ac{TSL} limit for each separation network and dataset. \autoref{sec:ss_dm} explores splitting signals to generate more train examples and whether \ac{DM} mitigates the effects gains found using \ac{TSL} limits. Final conclusions are provided in \autoref{sec:conclusions}.

\section{Signal Model}\label{sec:sigmod}
The noisy reverberant speech separation problem is defined as aiming to estimate $C$ speech signals $\hat{s}_c[i]$ for sample index $i$ and speaker number $c\in\{1,\ldots,C\}$ from the discrete time-domain mixture
\begin{equation}
    x[i] = \sum_{c=1}^{C} s_c[i] \ast h_c [i] + \nu [i]
\end{equation}
of length $L_x$. The $\ast$ operator denotes convolution, $h_c[i]$ is a \ac{RIR} corresponding to speaker $c$ and $\nu[i]$ denotes additive background noise.

\section{Separation Models}\label{sec:sepmodels}
\begin{figure}[!ht]
    \centering
    \includegraphics{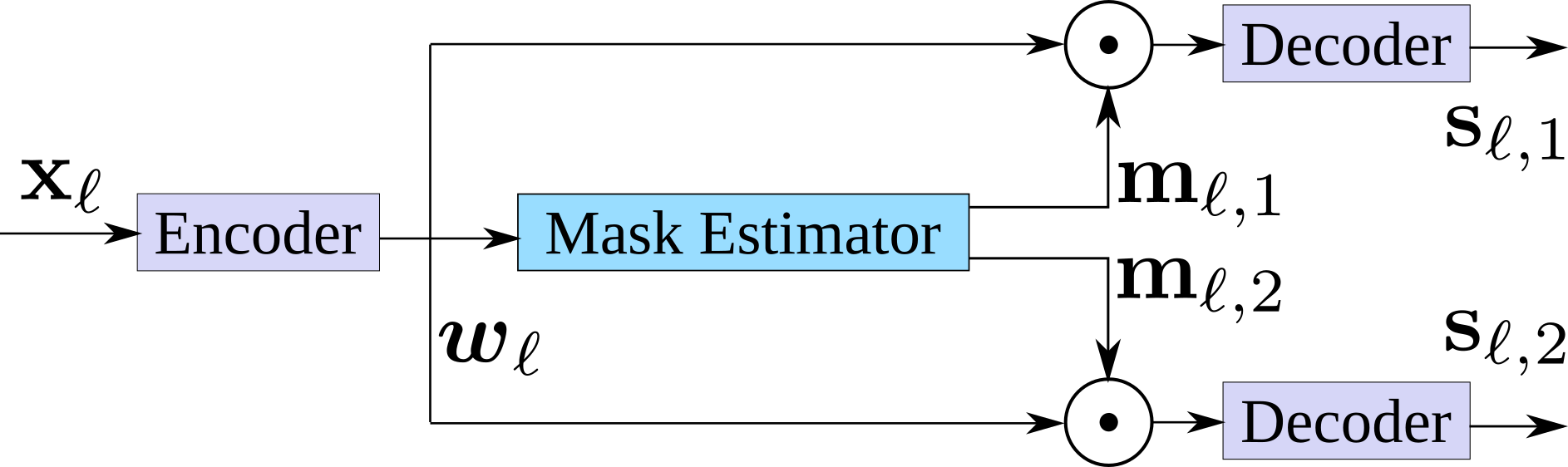}
    \caption{Architecture of the SepFormer and Conv-TasNet models, exemplary for $C=2$ speakers. The $\odot$ symbol denotes the Hadamard product.}
    \label{fig:tasnet}
\end{figure}

The SepFormer \cite{sepformer} and Conv-TasNet \cite{convtasnet} models are both widely researched \acp{TasNet}. The structure for these \acp{TasNet} is to have a time-domain neural encoder which encodes a mixture signal block $\vek{x}_\ell$ of size $L_\mathrm{BL}$  to $\vek{w}_\ell$ followed by a mask estimation network which estimates a series of masks $\vek{m}_\ell$ for each of $C$ speakers.
These masks are used to separate the encoded features which are then decoded back in to a time domain signal $\vek{s}_{\ell,c}$ using a neural decoder. An example of the architecture for $C=2$ speakers can be seen in \autoref{fig:tasnet}.
Both Conv-TasNet and SepFormer are trained using the utterance-level permutation-invariant \ac{SISDR} objective function \cite{upit,convtasnet,sepformer,LeRoux}. Models are trained according to the best performing models in each of their original papers \cite{convtasnet,sepformer} unless stated otherwise. Batch sizes of $2$ and $4$ are used for SepFormer and Conv-TasNet respectively except where otherwise stated.

\subsection{SepFormer Network} \label{sec:sepformer}
The SepFormer model is briefly introduced in this section. SepFormer is chosen because it is a large transformer model that is among the state-of-the-art models on several speech separation benchmarks \cite{sepformer, sepformer2}.
The SepFormer uses a 1D convolutional layer for encoding the signal proceeded by a \ac{ReLU} activation function. The decoder is a single transposed 1D convolutional layer. The mask estimation network uses a dual-path structure \cite{dprnn} whereby a series of Transformer layers are stacked such that each alternating layer computes a multihead attention layer on either the local or global context of the sequence. The local processing is achieved by first splitting the input signal into overlapping chunks of a predetermined size $K$ turning a batched 3D tensor into a 4D tensor. The output of the local Transformer layer is a 4D tensor which is then reshaped by swapping the axes of encoded chunks and the number of chunks before being fed into the global Transformer layer. The final stage is to reconstruct a 3D tensor using the overlap-add method to produce $C$ sequences of masks. The encoded features are then masked and decoded back into the time domain.

\subsection{Conv-TasNet}\label{sec:convtasnet}
The Conv-TasNet model contrasts the SepFormer in that it is a much smaller model (25.8M vs. 3.5M parameters in the implementations used here) and the only global information it processes is the overall signal energy, whereas the SepFormer model has global access to all information in the input signal due to the transformer layers used. Conv-TasNet uses a \ac{TCN} sequence model for the mask estimator in \autoref{fig:tasnet}.
 The encoder and decoder of the network are the same as those used for the SepFormer model but with a different number of output channels in the configuration used in this paper. The mask estimation network is composed of a 1D \ac{P-Conv} bottleneck layer, a \ac{TCN} and a 1D \ac{P-Conv} projection layer with a \ac{ReLU} activation function to produce the sequence of masks. The \ac{TCN} is composed of a series of convolutional blocks consisting of \ac{P-Conv} and \ac{DS-Conv} layers with kernel size $P$.
 The convolutional blocks are configured in stacks of $X$ blocks with increasing dilation factors $f\in\{2^0,2^1,\ldots,2^{X-1}\}$. Stack are repeated $R$ times with the dilation factor reset at the start of each stack~\cite{convtasnet}. 

 \section{Datasets}\label{sec:data}
Three corpora of 2-speaker mixtures are analysed in this work. The trends demonstrated later in Sections \ref{sec:evaluations} and \ref{sec:ss_dm} for the 2-speaker scenario are assumed to generalize to higher $C$ values. For all corpora, the $8$kHz \textit{min} configuration is used. The \textit{min} configuration refers to mixtures being truncated to the shortest utterance in a mixture as opposed to padding shorter utterances to the longest utterances.

 \subsection{WSJ0-2Mix and WHAMR}
WSJ0-2Mix and WHAMR are both simulated 2-speaker corpora derived from the WSJ0 corpus \cite{Isik,WHAMR}. WSJ0-2Mix takes speech samples from WSJ0 and overlaps them at \acp{SSR} between $0$ and $5$~dB. WHAMR is a noisy reverberant extension of WSJ0-2Mix with noise from the WHAM \cite{WHAM} dataset mixed with the loudest speaker at \acp{SNR} between $-6$ and $3$~dB.

\subsection{Libri2Mix}
Libri2Mix is a simulated 2-speaker mixture corpus derived from the LibriSpeech and WHAM corpora \cite{librimix}.  Speech samples come from the LibriSpeech corpus \cite{librispeech} and noise samples come from the WHAM corpus. Instead of \acp{SSR}, LibriMix uses \ac{LUFS} measured in dB to set the loudness of speakers and noise in the mixtures. Speakers have a loudness between -25 and -33~\ac{LUFS} and noise has a loudness between $-38$ and $-30$~\ac{LUFS}. For training, the \textit{train-100} dataset was chosen as it has a very similar \ac{TSL} distribution to the alternate \textit{train-360} dataset but with less examples, meaning shorter training times.

\subsection{Signal Length Distributions}
\begin{figure}[!ht]
    \centering
    \includegraphics{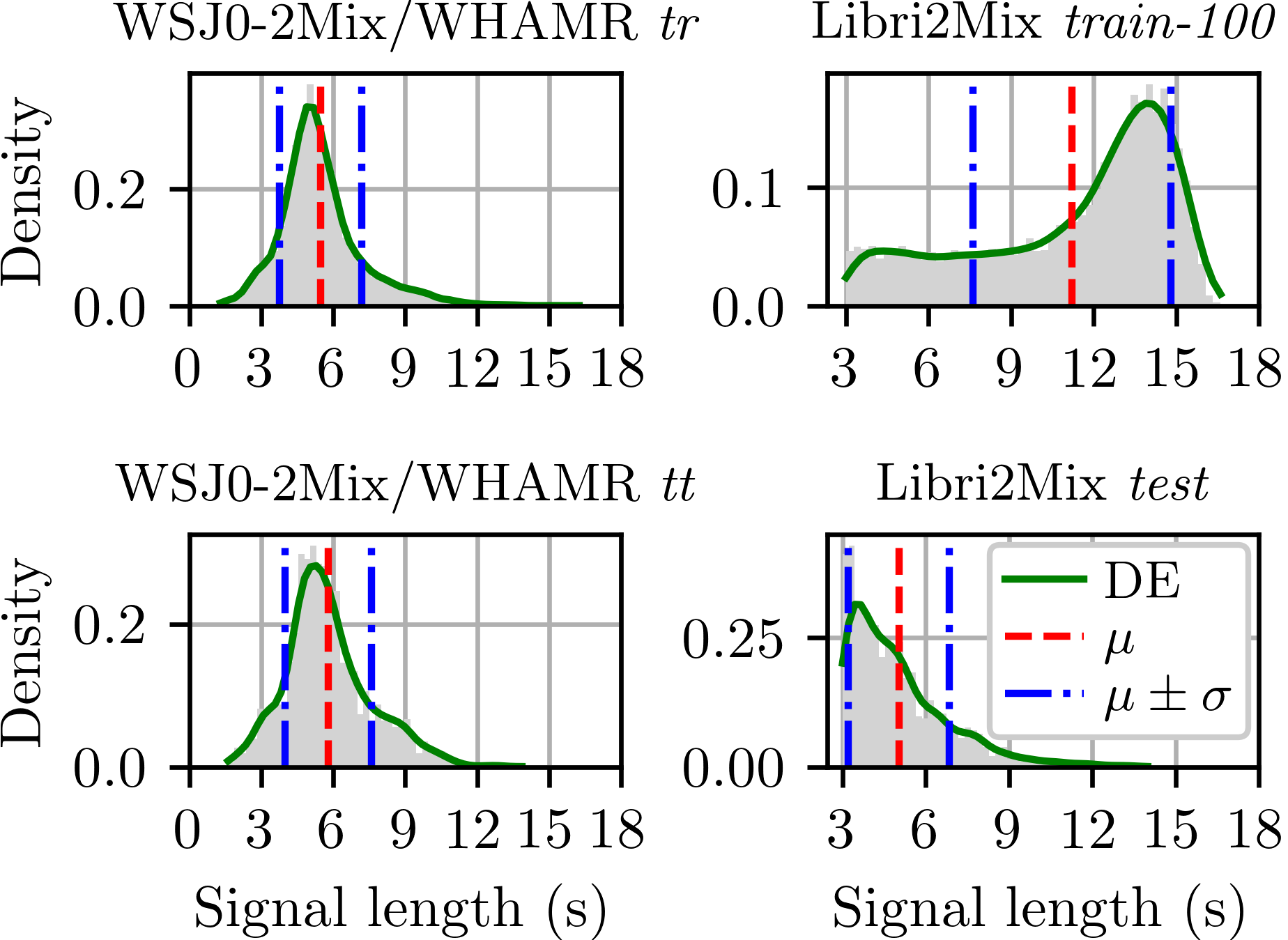}
    \caption{Distributions of mixture signal lengths in WSJ0-2Mix/WHAMR (left) and Libri2Mix (right) for both train (top) and test (bottom) sets. \Acf{DE} is shown by solid green lines, mean values are indicated by dashed red lines and standard deviation values by dash-dotted blue lines.}
    \label{fig:ddist}
\end{figure}

The distribution, \ac{DE} and the mean and standard deviation of the mixture signal length in the WHAMR train \textit{tr} and test \textit{tt} sets can be seen in \autoref{fig:ddist}. 
WSJ0-2Mix and WHAMR have identical signal distributions as WHAMR is derived from the former. These distributions are shown in the left panel. The train and test sets in WHAMR have similar distributions of signal length with mean values within 0.3s of one another and standard deviations within 0.1s of one another. This contrasts the distributions of the Libri2Mix dataset \cite{librimix} also shown in \autoref{fig:ddist} where the \textit{train-100} and \textit{test} sets have a difference in mean value of $6.2$s and difference in standard deviation of $1.79$s.

\section{Training Signal Length Analysis} \label{sec:evaluations}
Evaluations of varying the \ac{TSL} limit are presented in this section. For all evaluations, the improvement in \ac{SISDR} over the input mixture signal, denoted by $\Delta$ SISDR, is used as the evaluation metric. \Ac{SISDR} measures the energy of distortions in the estimated speech signals and is one of the most common metrics used in recent monaural speech separation literature \cite{dprnn,sepformer,QDPN}.

\subsection{Initial TSL Limit Evaluations}\label{sec:corporaeval}
\begin{figure}[!ht]
    \centering
    \includegraphics{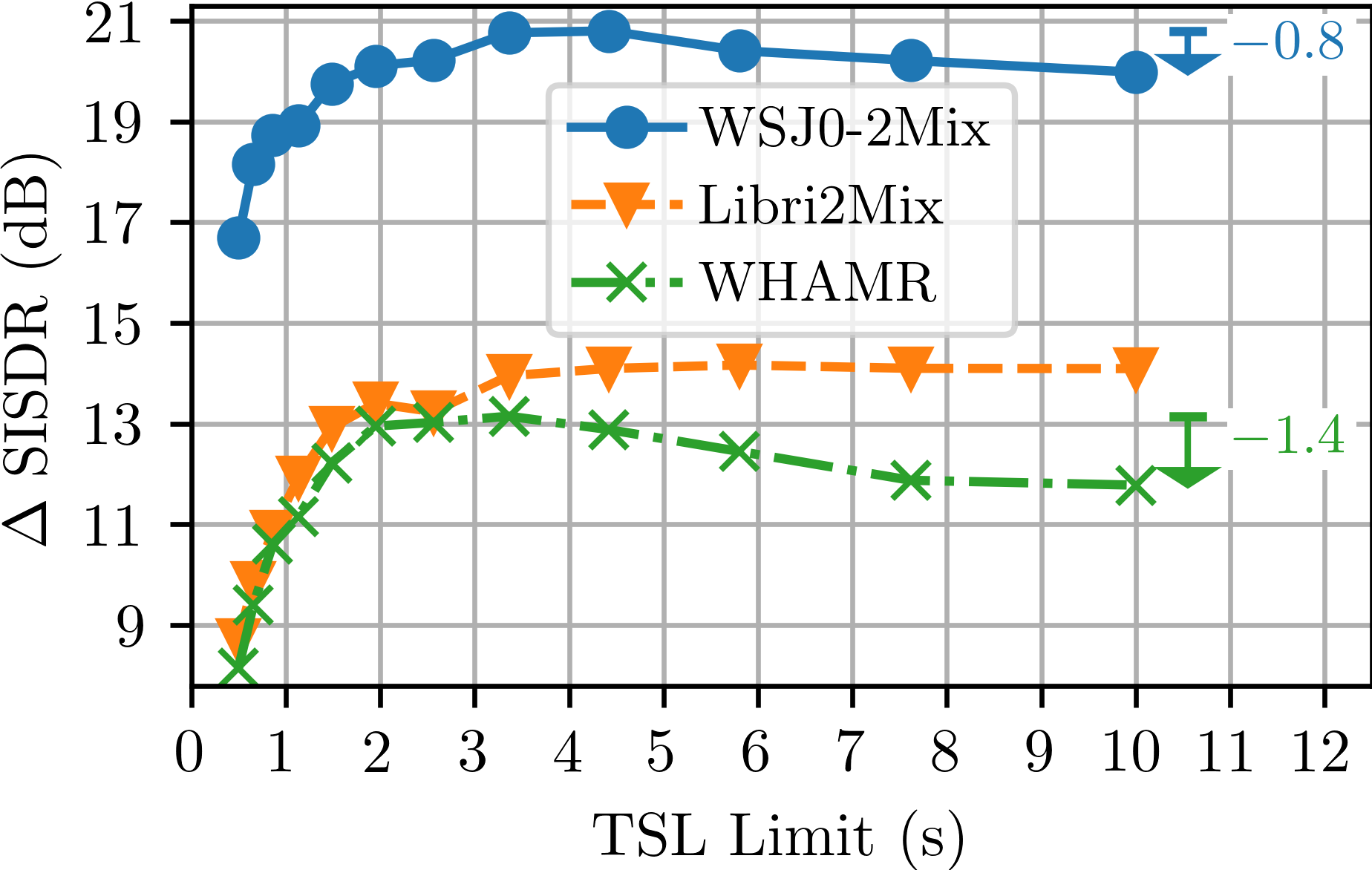}
    \caption{SepFormer results for varying the \ac{TSL} limit for the anechoic WSJ0-2Mix (top), Libri2Mix (middle) and WHAMR (bottom) test sets.}
    \label{fig:corporacomp}
\end{figure}

As a first experiment, twelve SepFormer models are trained and evaluated on WSJ0-2Mix, WHAMR and Libri2Mix, each with a different \ac{TSL} limit. Twelve logarithmically spaced signal limits $T_\mathrm{lim}$ are selected between $0.5$s and $10$s:
\begin{align}
    T_\mathrm{lim} \in \{&0.5, 0.66, 0.86, 1.13, 1.49, 1.95,  \nonumber \\
    &2.56, 3.36, 4.42, 5.8, 7.62, 10\}\text{s}.
\end{align}
The notation $L_\mathrm{lim}$ is used for the respective discrete sample index, i.e.~$L_\mathrm{lim}=T_\mathrm{lim}\mathrm{f}_s$ for sampling rate $\mathrm{f}_s$.
When cutting the training signal lengths such that $L_x \le L_\mathrm{lim}$ the starting sample index of the signal is randomly selected from the uniform distribution $\mathcal{U}\left(0,1+max\left(0,L_x-L_\mathrm{lim}\right)\right)$.
Performance for SepFormer models trained and evaluated on all three datasets is compared in \autoref{fig:corporacomp}. 
For the WHAMR corpus, an increase in overall $\Delta$ \ac{SISDR} performance from the $0.5$s to $1.95$s limit can be observed. The optimal \ac{TSL} is at $3.36$s. Between $3.36$s and $10$s performance decreases again by $-1.4$dB. This may seem surprising as the general convention with training \acp{DNN} is that more data normally results in improved overall performance. A similar trend is observed for WSJ0-2Mix where there is a notable increase between 0.5s and $3.36$s and then a drop in performance of $0.8$dB between $4.4$s and $10$s. For Libri2Mix, the performance saturates before a \ac{TSL} limit of $4.42$s. There is no drop in performance as the \ac{TSL} limit approaches $10$s which is likely due to the Libri2Mix training set having a more uniform distribution below signal lengths of $10s$ than the WHAMR or WSJ0-2Mix datasets, cf.~\autoref{fig:ddist}.

\begin{figure}[!ht]
    \centering
    \includegraphics{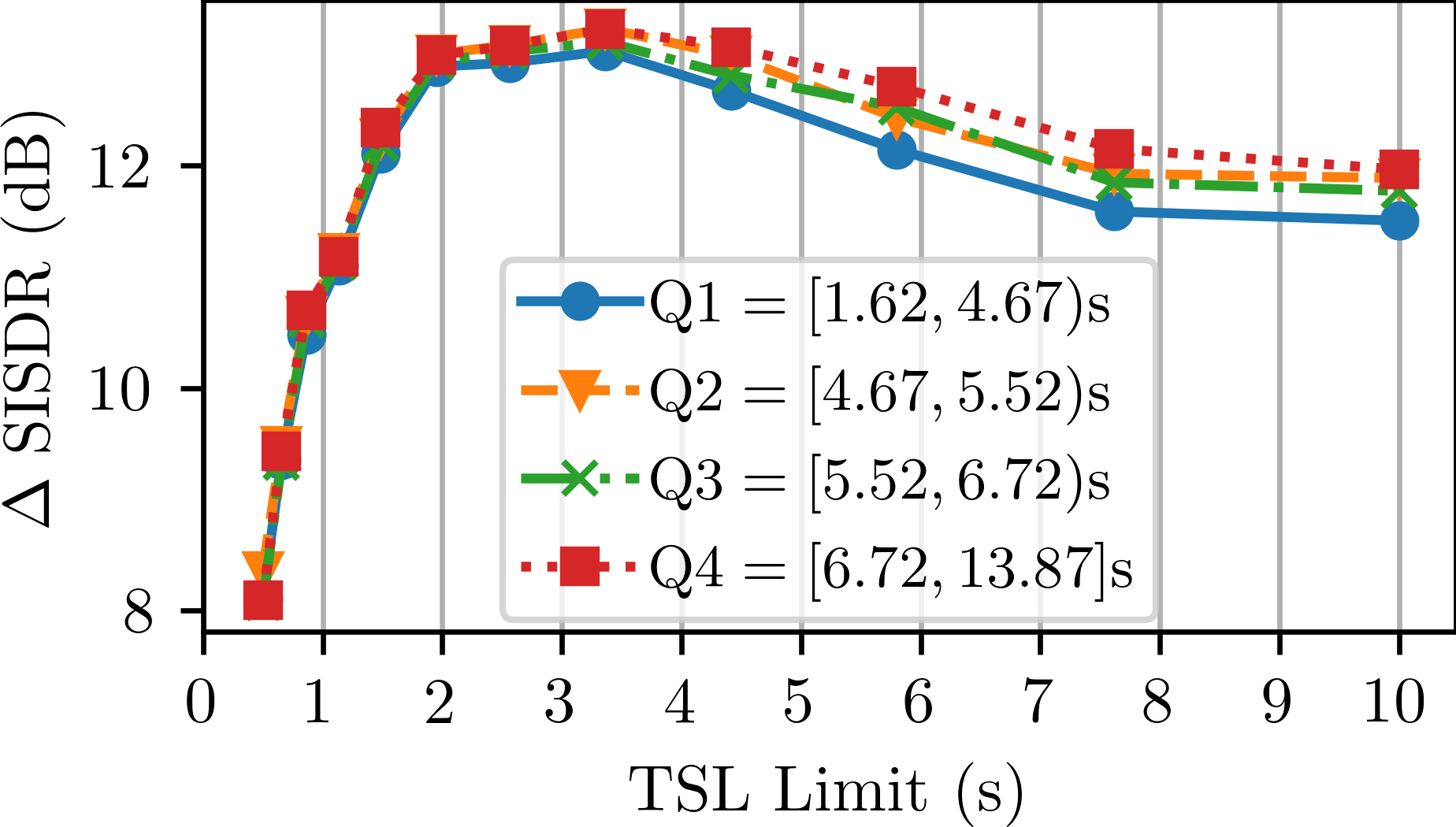}
    \caption{\Acf{TSL} analysis of the $1$st to $4$th signal length quartiles in the WHAMR evaluation set.}
    \label{fig:quartiles}
\end{figure}

The results for the WHAMR evaluation set are separated into quartiles of mixture signal length for the following experiment. 
$\Delta$ \ac{SISDR} results for each quartile are shown in \autoref{fig:quartiles}. 
Comparing Q1 to Q4 shows that, with a sufficiently large \ac{TSL} ($\ge 1.95$s), the best separation performance in \ac{SISDR} is found on the longest signal lengths, regardless of \ac{TSL}. A loss in \ac{SISDR} performance is still observed from $3.36$s to 10s regardless of which quartile is evaluated. 

\subsection{Training Time Evaluation}
\begin{figure}[!ht]
    \centering
    \includegraphics{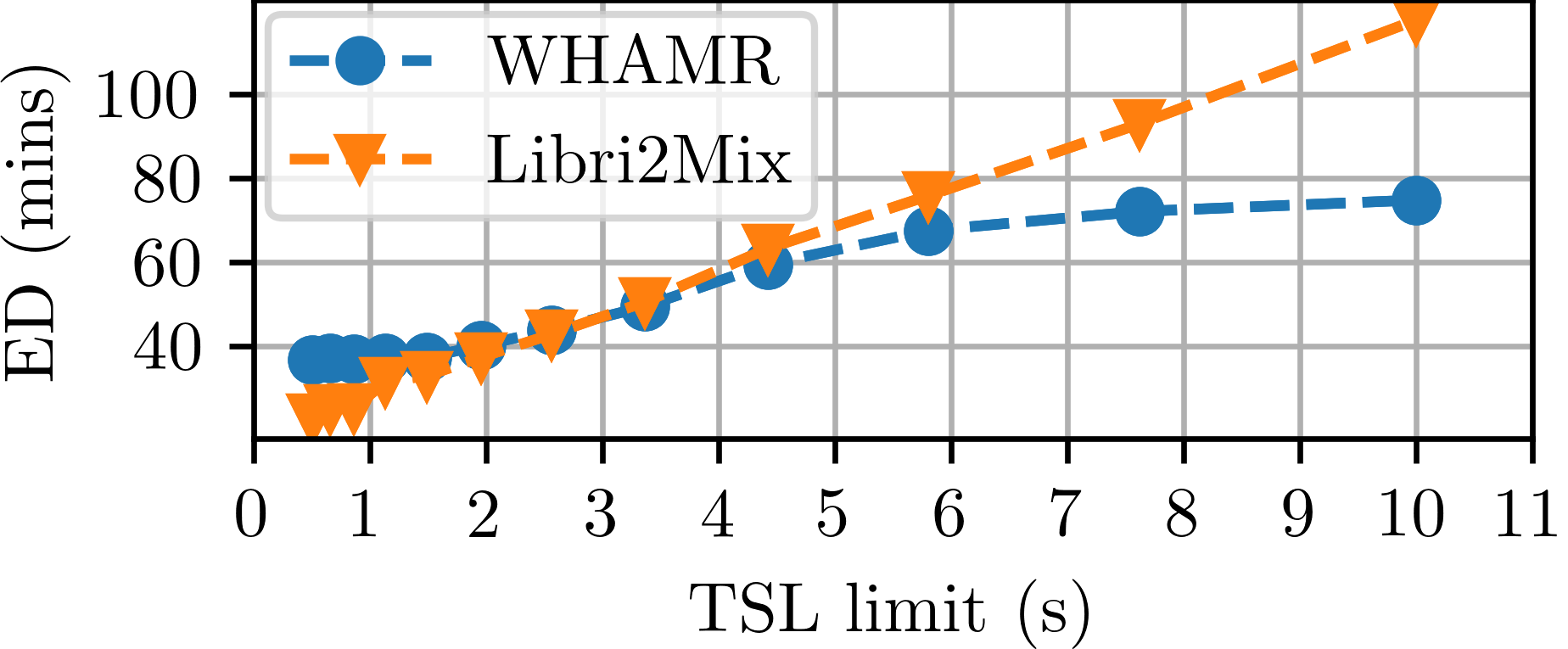}
    \caption{Comparison of average epoch duration (in mins) for the SepFormer model on the Libri2Mix and WHAMR training sets.}
    \label{fig:epochtime}
\end{figure}

The average training \ac{ED} for the SepFormer model on WHAMR and Libri2Mix training sets are shown in \autoref{fig:epochtime}. Note the \ac{ED} for WSJ0-2Mix is omitted for brevity but is similar to WHAMR due to having the same \ac{TSL} distribution (cf.~\autoref{fig:ddist}). All models were trained on the same hardware to control any impact this has on speed.
The average \acp{ED} for the WHAMR dataset have a sigmoidal shape due to the majority of the signal lengths being concentrated around the mean signal length of the training set ($5.6$s, cf.~\autoref{fig:ddist}). Libri2Mix has a more linear relationship between \ac{TSL} limit and \ac{ED} due to the more uniform shape of the signal length distribution below $10$s in the \textit{train-100} set, cf.~\autoref{fig:ddist}. Reducing the \ac{TSL} limit has more benefit in terms of \ac{ED} for the Libri2Mix dataset when iterating over all training examples.

\subsection{Fixed vs. Random Start Index}\label{sec:fixedvsrand}
\begin{figure}[!ht]
    \centering
    \includegraphics{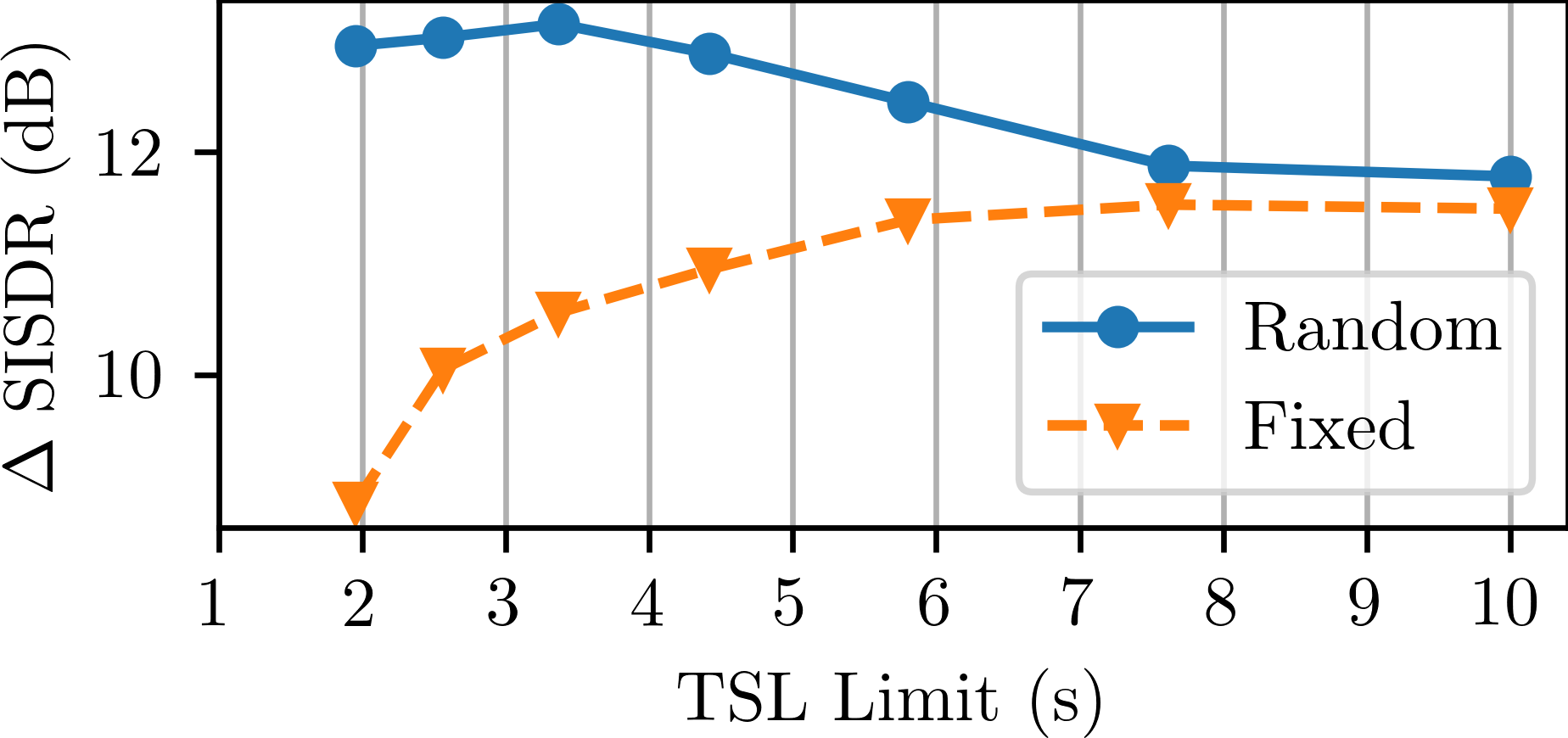}
    \caption{Comparison of \ac{TSL} variation for the SepFormer model trained and evaluated on the WHAMR datasets on a subset of \ac{TSL} limits in the range $[1.95, 7.62]$s.}
    \label{fig:fixedvsrand}
\end{figure}

In Section \ref{sec:corporaeval}, the start index of each shortened signal was randomly sampled from a uniform distribution. In this section, this is compared to using a fixed start sample. A start sample of 1999 ($=0.25$s at 8kHz) was used for signals where the original mixture signal length was larger than $L_\mathrm{lim}$, else the entire signal length was used. The motivation for this was that many training examples contain silence at the beginning of clips. It was considered desirable to omit as much silence to make for a fairer comparison with the randomly sampled clips which are assumed to have a lower likelihood of 
 beginning with silence.
Results in \autoref{fig:fixedvsrand} confirm that the loss in performance from a \ac{TSL} limit of $3.36$s to $10$s with WHAMR is due to randomly sampling the start index.
The performance saturates at a \ac{TSL} limit of $5.8$s when using a fixed start index. This is similar to the performance saturation point of Libri2Mix in \autoref{fig:corporacomp}  demonstrating the performance drop in higher \acp{TSL} seen before on WHAMR (cf.~\autoref{fig:corporacomp}) is related to both a non-uniform \ac{TSL} distribution and the random sampling used.

\subsection{Transformer vs. Convolutional Model}
\begin{figure}[!ht]
    \centering
    \includegraphics{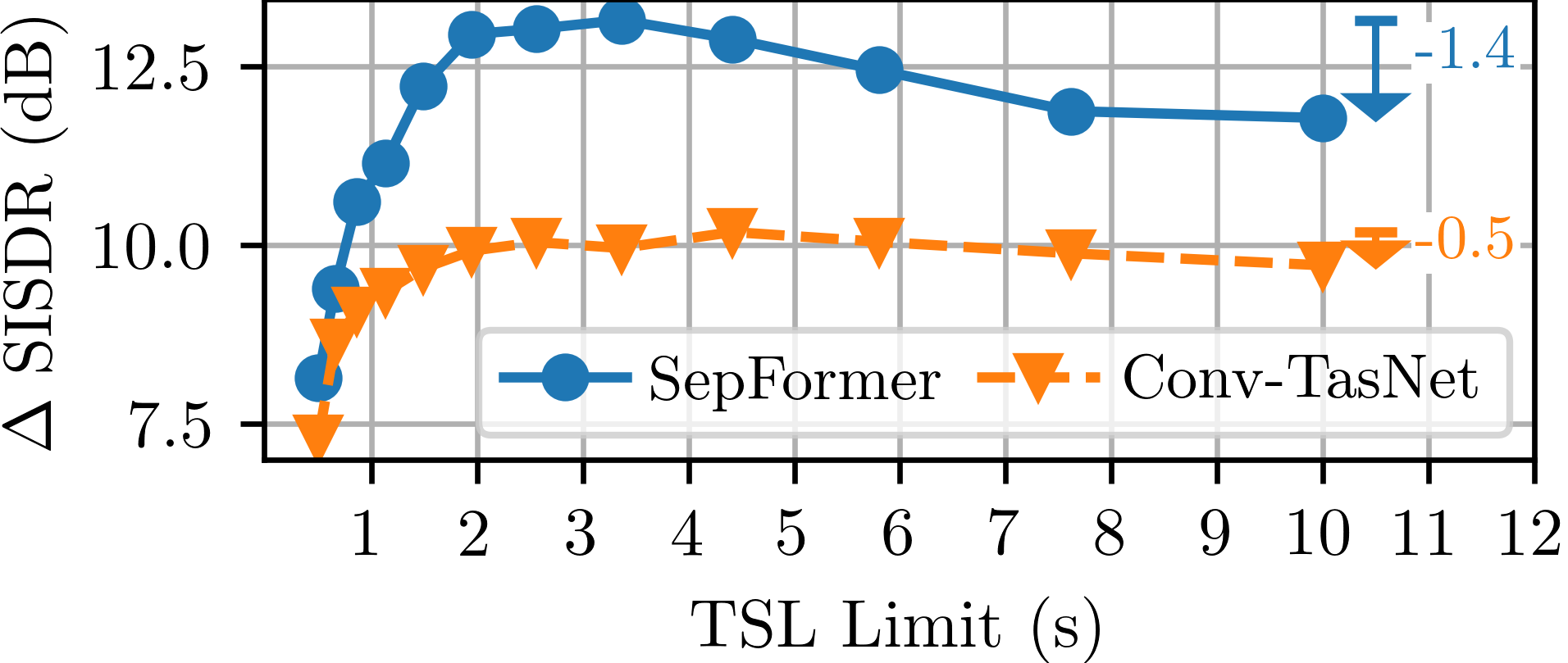}
    \caption{Comparison of SepFormer and Conv-TasNet across \ac{TSL} limits $T_\mathrm{lim}\in [4.42,7.62]$ using the WHAMR corpus.}
    \label{fig:modelcomp}
\end{figure}

Results comparing the Conv-TasNet model (cf.~\autoref{sec:convtasnet}) to the SepFormer model are shown in \autoref{fig:modelcomp}. The loss in performance above $3.36$s is not observed for the Conv-TasNet model. All \ac{SISDR} results above $T_\mathrm{lim}=1.95$s are within in $0.5$dB of each other suggesting Conv-TasNet is more invariant to the \ac{TSL} limit if the limit is sufficiently large. This is possibly due to the $1.53$s receptive field of the Conv-TasNet models being smaller than these particular \acp{TSL} limits~\cite{rfield}.

\section{Signal Splitting and Dynamic Mixing}\label{sec:ss_dm}
Next, two sampling strategies are evaluated to (i) investigate whether the performance gained by \ac{TSL} with random sampling on shorter sequences still holds when the same quantity of audio data in terms of length in seconds is used and (ii)  whether using \ac{TSL} limits still result in performance gains with \ac{DM}, i.e. simulating new speech mixtures for each epoch \cite{wavesplit}. 

\subsection{Signal Splitting}
\begin{figure}[!ht]
    \centering
    \includegraphics{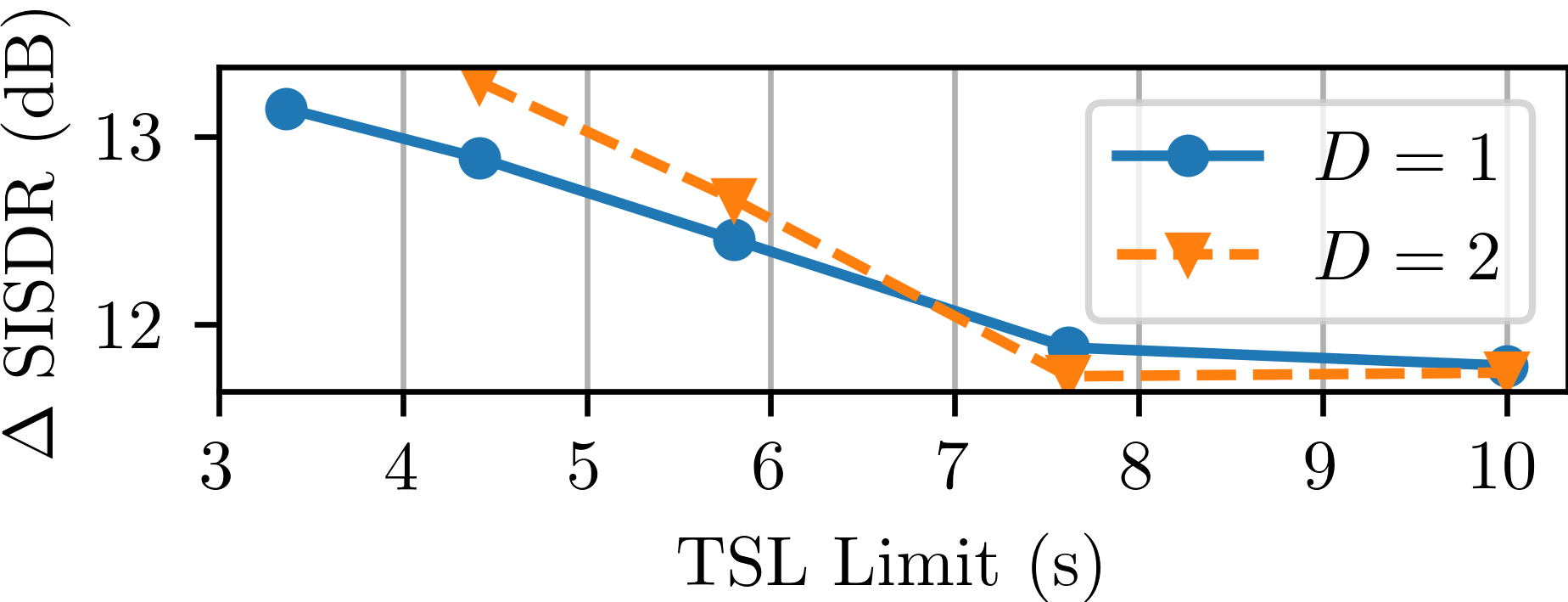}
    \caption{Comparison of split signal and batch reshape training $D=2$ against full signal training $D=1$ for the SepFormer model.}
    \label{fig:split_signal}
\end{figure}

A signal splitting strategy was designed such that a batch of inputs $\vekt{X}\in \Real^{M \times L_x}$ was reshaped to $\vekt{X}'\in \Real^{MD\times\frac{L_x}{D}}$ for batch size $M$. Signal length $L_x$ is still limited such that $L_x\le L_\mathrm{lim}$. The motivation of this method is to evaluate the importance of training on the entire sequence length compared to the raw data quantity used in seconds. Computational complexity in training is also reduced. \acp{TSL} for $T_\mathrm{lim} \in [4.42,10]$s are analysed for $D=2$.
\autoref{fig:split_signal} shows that $D=2$ improves performance for shorter \ac{TSL} limits ($T_\mathrm{lim} \le 5.8$s) compared to $D=1$ (the original shape). However for $T_\mathrm{lim} \ge 7.62$s the performance is similar to $D=1$. As in \autoref{fig:corporacomp} this is likely due to the \ac{TSL} distribution of WHAMR.

\subsection{Dynamic Mixing}
\begin{figure}[!ht]
    \centering
    \includegraphics{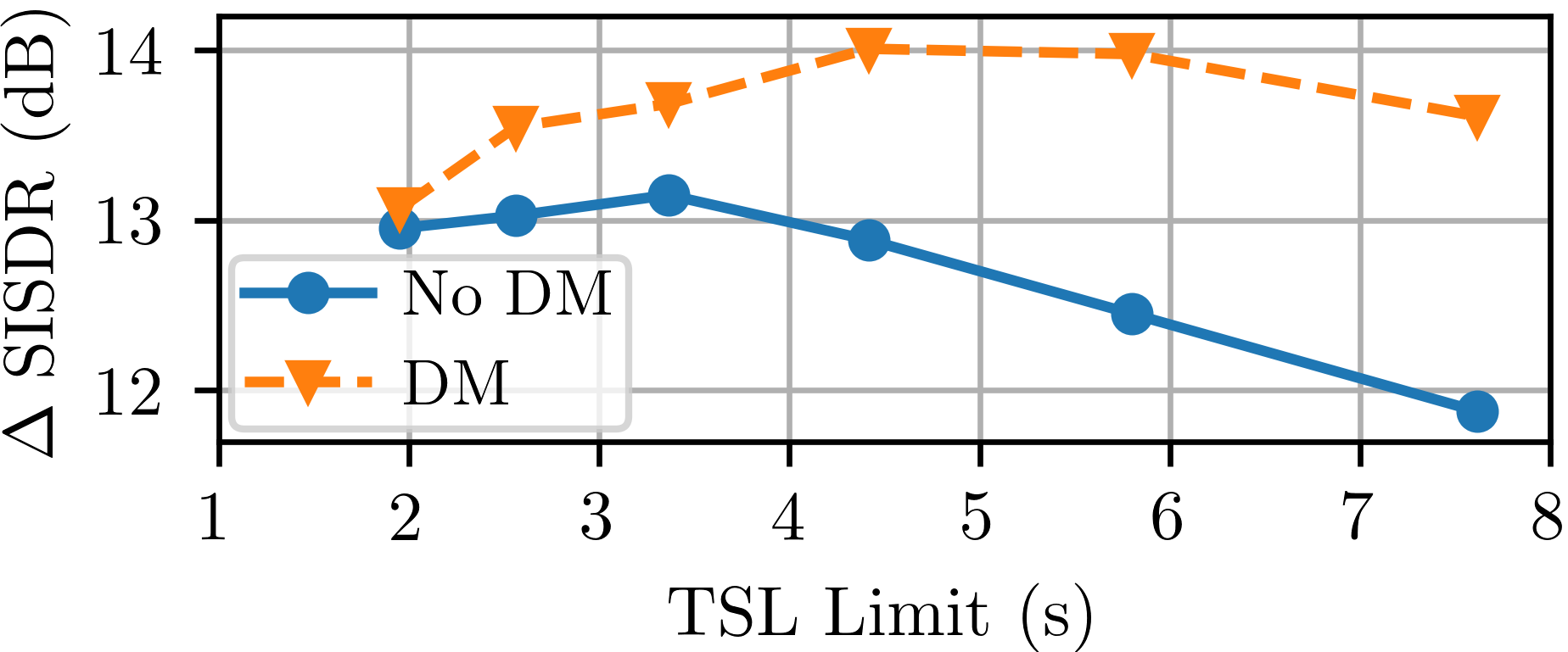}
    \caption{Comparison of SepFormer model training using \ac{TSL} limits with and without dynamic mixing being used as well on WHAMR evaluation set.}
    \label{fig:no_dmvsdm}
\end{figure}

\Ac{DM} was proposed to improve performance of various speech separation models 
 \cite{wavesplit}. \ac{DM} often results in a $1.0$ to $1.5$dB  \ac{SISDR} performance improvement dependant upon the model and dataset \cite{wavesplit,  deformtcn}. The random start index sampling used in \autoref{sec:evaluations} is similar to \ac{DM} in that it provides the model with unique training examples each epoch but without simulating new mixtures. In this section using \ac{DM} and \ac{TSL} limits is compared against just using \ac{TSL} limits to see if further performance gains can be attained using both approaches.
The \ac{DM} results for the WHAMR corpus are shown in \autoref{fig:no_dmvsdm}. It can be seen that with  \ac{DM} the drop in performance is less (at $7.62$s) than without \ac{DM}.
The best performing model $T_\mathrm{lim}=4.4$s is compared to the Sepformer model with no TSL limit in \autoref{tab:dmvsrs}. The batch size reported for the model with no TSL is the largest it was found possible to train on a $32$GB Nvidia V$100$ GPU. It can be seen that using the TSL limited model is able to match its performance with an average \ac{ED} reduction of $44$\%, highlighting the benefit of this approach.
\begin{table}[!ht]
\setlength\tabcolsep{2.5pt}
\centering
\caption{Comparison of best performing SepFormer models on WHAMR with and without \ac{TSL} limits. $M$ denotes batch size and the average \acf{ED} is reported in minutes.}
\begin{tabular}{|c|c|c|c|c|}
\hline
\rowcolor[HTML]{C0C0C0} 
\textbf{Model} & $M$ & $T_\mathrm{lim}$ \textbf{(s)} & \textbf{$\Delta$ SISDR (dB)} & \textbf{ED (mins)} \\ \hline
SepFormer + DM & 1   & $- $                 & \textbf{14.0}                  & 85                 \\
SepFormer + DM & 2   & 7.62                 & 13.6                  & 74                 \\
SepFormer + DM & 2   & 4.42                 & \textbf{14.0}                  & \textbf{59}        \\ \hline
\end{tabular}
\label{tab:dmvsrs}
\end{table}

\section{Conclusion}\label{sec:conclusions}
In this paper it was shown that \ac{TSL} limits can affect the overall performance for speech separation models in a number of ways. For WSJ0 derived speech separation benchmarks, i.e.~WSJ0-2Mix and WHAMR, it is optimal to use shortened training examples randomly sampled from the original examples due to the signal length distribution of these corpora. For the Libri2Mix dataset, the same method led to shorter training times with no notable loss in performance.
The SepFormer model was compared to the Conv-TasNet model and it was shown that the Conv-TasNet performance has less variation. Using dynamic mixing and TSL limits with random sampling was shown to be able to match the performance of the SepFormer model trained \ac{DM} using full sequence lengths on WHAMR with a 44\% reduction in training time. 
With some previous literature opting to limit \acp{TSL} \cite{convtasnet,dprnn} and others not \cite{QDPN,sepformer} on the same benchmarks, the results in this paper suggest that this is not a fair comparison and that \ac{TSL} limiting is important to factor in when analysing results, particularly for the WSJ0-2Mix and WHAMR benchmarks.

\bibliographystyle{IEEEtran}
\bibliography{refs}

\end{document}